\documentclass[usenatbib,useAMS]{mn2e}
\usepackage{epsfig,color}

\newcommand{\gcc}{\ \mathrm{g\ cm^{-3} }}
\newcommand{\nuc}[2]{\ensuremath{\mathrm{^{#1}#2}}}

\newcommand{\msun}{\ensuremath{M_\odot}}
\newcommand{\simgt}{\,\hbox{\lower0.6ex\hbox{$\sim$}\llap{\raise0.6ex\hbox{$>$}}}\,}
\title[Tracer particle method resolution study]{Nucleosynthesis in
  thermonuclear supernovae with tracers: convergence and
  variable mass particles}

\author[Seitenzahl et al. 2010]{I.~R.~Seitenzahl$^{1}$,
  F.~K.~R\"opke$^{1}$, M.~Fink$^{1}$, R.~Pakmor$^{1}$\\
$^{1}$Max-Planck-Institut f\"ur Astrophysik,
                 85741 Garching, Germany\\} 
\date{\today}
\pubyear{}
\begin{document}
\maketitle
\begin{abstract}
  Nucleosynthetic yield predictions for multi-dimensional simulations
  of thermonuclear supernovae generally rely on the tracer particle
  method to obtain isotopic information of the ejected material for a
  given supernova simulation.  We investigate how many tracer
  particles are required to determine converged integrated total
  nucleosynthetic yields.  For this purpose, we conduct a resolution
  study in the number of tracer particles for different hydrodynamical
  explosion models at fixed spatial resolution.  We perform
  hydrodynamic simulations on a co-expanding Eulerian grid in two
  dimensions assuming rotational symmetry for both pure deflagration
  and delayed detonation Type Ia supernova explosions.  Within a given
  explosion model, we vary the number of tracer particles to determine
  the minimum needed for the method to give a robust prediction of the
  integrated yields of the most abundant nuclides.  For the first
  time, we relax the usual assumption of constant tracer particle mass
  and introduce a radially varying distribution of tracer particle
  masses.  We find that the nucleosynthetic yields of the most
  abundant species (mass fraction $ > 10^{-5}$) are reasonably well
  predicted for a tracer number as small as 32 per axis and direction
  -- more or less independent of the explosion model.  We conclude
  that the number of tracer particles that were used in extant
  published works appear to have been sufficient as far as integrated
  yields are concerned for the most copiously produced nuclides.
  Additionally we find that a suitably chosen tracer mass distribution
  can improve convergence for nuclei produced in the outer layer of
  the supernova where the constant tracer mass prescription suffers
  from poor spatial resolution.
\end{abstract}

\begin{keywords}{nuclear reactions, nucleosynthesis, abundances --
    supernovae: general}
\end{keywords}

\section{Introduction}
\label{sec:int}
The basic processes of Type~Ia supernovae (SNe~Ia) have been proposed
almost 50 years ago \citep[see e.g.][for a review]{hillebrandt2000a}:
A thermonuclear explosion in electron-degenerate matter
\citep{hoyle1960a} produces radioactive \nuc{56}{Ni} that by its decay
delivers energy at exactly the rate observed in SN~Ia light curves
\citep{pankey1962a,truran1967a, colgate1969a, kuchner1994a}.  Despite
this long history, the questions of the progenitor system and the
explosion scenario are not completely answered.  Pure deflagrations in
Chandrasekhar-mass white dwarfs, currently somewhat disfavoured for
their apparent inability to produce bright explosions, have been one
of the contending explosion scenarios for a long time
\citep[e.g.][]{nomoto1984a,gamezo2003a,roepke2007c}.  If the initial
deflagration can transition into a detonation
\citep[e.g.][]{khokhlov1997a,roepke2007d,woosley2007a,woosley2009a},
then better agreement of the models with observations can be obtained
\citep[e.g.][]{roepke2007b,bravo2008a,kasen2009a}.  If such a
deflagration to detonation transition does not occur \citep[cf.\
e.g.][]{niemeyer1999a}, the possibility that off-center ignition in a
single spot leads to a detonation of a white dwarf out of hydrostatic
equilibrium in the so-called GCD model is yet another proposed
mechanism \citep{plewa2004a,jordan2008a,meakin2009a}, although the
robustness of the initiation of the detonation in this model is also
far from certain \citep[cf.][]{roepke2007a, seitenzahl2009c,
  seitenzahl2009b}.  However, there are indications from recent
stellar population synthesis studies
\citep[e.g.][]{ruiter2009a,mennekens2010a} and X-ray observations of
elliptical galaxies and galaxy bulges \citep{gilfanov2010a}, that the
long favoured single degenerate Chandrasekhar-mass progenitor channel
is unable to account for observational rates of SNe Ia
\citep{cappellaro1999a}.  Supernovae resulting from the merger of two
white dwarfs \citep[e.g.][]{pakmor2010a} have more favourable
statistics and remain a possible explosion channel.  Last but not
least, the double detonation sub-Chandrasekhar mass model, which has
received renewed interest of late \citep{fink2007a,fink2010a,
  sim2010a}, is currently again considered a serious alternative.

The different nucleosynthesis occurring in all these explosion
scenarios is an important test for the validity of the respective
models.  For one-dimensional simulations, the nucleosynthesis can be
calculated during the simulation via the coupling of a nuclear
reaction network to the hydrodynamics. Since today's most promising
explosion models either explicitly break spherical symmetry (e.g.\
mergers, double detonations, off-center ignitions) or, in the case of
centrally ignited spherically symmetric explosions include buoyancy
driven turbulent combustion, at least two-dimensional simulations are
required to simulate the essential physics of the explosion.
Unfortunately, in highly resolved two-dimensional simulations and
especially in three-dimensional simulations use of an extended, full
nuclear reaction network during the hydrodynamic evolution is
computationally not feasible.  Consequently, the nuclear energy
release for multi-dimensional simulations is modeled with simplified
and approximate schemes
\citep[e.g.][]{khokhlov1995a,reinecke2002b,vladimirova2006a,calder2007a,townsley2007a,seitenzahl2009a}.
Although some works on multi-dimensional explosion simulations of SN
Ia do not calculate any detailed nucleosynthesis at all
\citep[e.g.][]{reinecke2002d,plewa2004a,gamezo2005a,plewa2007a,jordan2008a,bravo2008a},
a detailed isotopic composition of the ejecta is often determined in a
post-processing step
\citep[][]{travaglio2004a,travaglio2005a,brown2005a,roepke2006b,fink2010a,maeda2010a}.
While alternative methods for obtaining isotopic nucleosynthetic
yields for SNe Ia are currently under investigation
\citep{seitenzahl2008b,meakin2009a}, the tracer particle method still
remains currently the only viable choice for the task.  In
Section~\ref{sec:tpm} we briefly review the tracer particle method and
introduce how we distribute tracer particles of variable mass in our
initial models.  In Section~\ref{sec:sim} we give details about our
hydrodynamic simulations.  Section~\ref{sec:conv} presents the results
of a convergence study in the number of tracer particles and
highlights the advantages of using a radially varying distribution of
tracer masses.  We conclude the paper with a summary and discussion in
Section~\ref{sec:conc}.

\section{Tracer particle method}
\label{sec:tpm}
Tracer particles are a Lagrangian component in an Eulerian grid code.
The particles are assigned masses and positioned in such a way that a
density profile reconstructed from their distribution resembles that
of the underlying star.  During the hydrodynamical simulation, they
are advected by the flow, recording the history of thermodynamic
conditions along their path.  The tracers are nevertheless considered
\emph{massless} in the sense that the mass they represent does not
couple to the hydrodynamic flow via gravity or inertia -- they are
simply passively advected by the flow along streamlines.

The tracer particle method in multiple dimensions was employed first
by \citet{nagataki1997a} in the context of core collapse supernovae
(note that \citet{thielemann1986a} already post-processed Lagrangian
mass zones of 1D hydrodynamical calculations of SNe Ia).

\subsection{Tracer particle masses and placement}
\label{sec:placement}
Historically, $N$ tracer particles are placed in a star of radius $R$
and mass $M$, such that each tracer particle represents the same
amount of mass $m=M/N$.  However, other choices for the distribution
of tracer masses are possible and may be sometimes preferable.  In
this paper we consider a tracer mass distribution that varies
\emph{smoothly} with radius and additionally fulfills the following
criteria:
\begin{eqnarray}
  \label{eq:dist}
  \textrm{for}\qquad \quad 0 < &r& < R_1 \; : \quad \textrm{const.\ mass
    per particle} \nonumber \\ 
  \textrm{for}\qquad \; R_1 < &r& < R_2 \; : \quad
  \textrm{const.\ volume per particle} \\
  \textrm{for}\qquad \; R_2 < &r& < R \;\;\: : \quad
  \textrm{const.\ mass per particle,} \nonumber
\end{eqnarray}
where $0 < R_1 < R_2 < R$. \\

\noindent This choice of the functional form of the radial
distribution of the tracer particles is motivated by the following
considerations: in the inner part of the star ($0<r<R_1$) the burning
occurs at high density and tracer particles of constant mass are
sufficiently spatially dense to resolve the abundance gradients. In a
transition zone ($R_1<r<R_2$), where the density is lower and nuclear
burning is incomplete, intermediate mass elements are synthesized.
Due to the relatively low density, a constant tracer mass approach
results in a small spatial density of particles there. The constant
volume requirement effectively moves tracers from the well sampled
inner regions to regions of lower initial density where interesting
nucleosynthesis occurs. Finally, a constant volume approach all the
way out to the surface of the star would result in a wastefully large
amount of tracer particles placed into the very outer layers at very
low density where usually no nucleosynthesis occurs. The transition
back to constant tracer masses in the very outer layers where hardly
any nucleosynthesis occurs ($R_2<r<R$) alleviates this problem.

In the case of equally massive tracer particles, the white dwarf is
divided into $N_r$ shells of equal mass. Then the same number of
tracers, $N_{\theta}$, is distributed uniformly in mass into each of
these shells. If the tracers should represent the same volume, the
shells are chosen to have the same volume instead. The distribution of
the tracers in these shells is done uniformly in mass as before. If
both approaches are mixed as described above, we require in addition
that the tracers on either side of the interface between the areas of
equal-mass and equal-volume all have the same mass. Together with the
constraints~(\ref{eq:dist}), this leads to the following linear
equations constraining the particle numbers $N_1,N_2,N_3$ in each of
the three regions:
\begin{eqnarray}
  \label{eq:linsys}
  \frac{M_1}{N_1}&=& \frac{V \rho_1}{N_2} \nonumber \\ 
  \frac{M-M_2}{N_3} &=& \frac{V \rho_2}{N_2} \\ 
  N_1+N_2+N_3 &=& N_r \;, \nonumber 
\end{eqnarray}
where $M_i=M(R_i)$ is the enclosed mass as a function of radius,
$\rho_i=\rho(R_i)$ is the mass density at position $R_i$
and $V=[4 \pi (R_2^3-R_1^3)]/3$.\\
Note that for a cold WD in hydrostatic equilibrium of known
composition and total mass $M$, given any one of the quantities
$\{R_i,M_i,\rho_i\}$ the other two are uniquely determined by the
equation of state.  

\noindent The solution to the set of linear equations
(\ref{eq:linsys}) is given by:
\begin{eqnarray}
  \label{eq:distN}
  N_1 &=& \frac{A}{A+B+1} N_r \nonumber \\ 
  N_2 &=& \frac{B}{A+B+1} N_r \\ 
  N_3 &=& \frac{1}{A+B+1} N_r \; ,\nonumber
\end{eqnarray}
where $A=M_1/(V \rho_1)$ and $B=(M-M_2)/(V \rho_2)$.

\begin{figure}
  \includegraphics[width=\columnwidth]{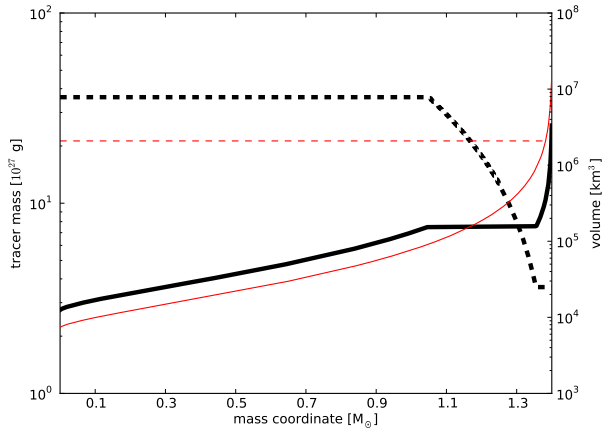}
  \caption{Shown are the mass (dashed lines) and volume (solid lines)
    represented by a tracer particle as a function of its initial
    position in the mass coordinate for $\mathrm{N}=256 \times 512$
    tracers. Thin (red) lines are for constant tracer particle
    mass. Thick (black) lines are for variable tracer particle mass
    for $M_1=1.05 \, \msun$ and $M_2=1.355 \, \msun$.}
  \label{fig:vol_mass}
\end{figure}

\noindent The tracers are placed into the star according to the rules
outlined above. An additional small offset is added to the coordinates
such that each particle has a random initial position within its
corresponding fluid element.  A comparison of the constant and
variable tracer mass distributions (for the same number of total
tracer particles) is shown in Fig.~\ref{fig:vol_mass}. It is evident
from the figure that, for this particular choice of $M_1=1.05 \,
\msun$ and $M_2=1.355\, \msun$, in the outer regions of the star
($M(r)/\msun \simgt 1.15$) the variable tracer mass approach results
in a smaller volume represented by each particle.  This increased
spatial resolution at low density comes at the expense of slightly
larger particle masses in the inner regions of the star.

\section{Simulations}
\label{sec:sim}
The code used to simulate the supernova explosions is the MPA SN Ia
code \citep[see, e.g.,][]{reinecke1999b,reinecke2002b}.  In this
Eulerian hydrodynamics code the reactive Euler equations are solved
using a finite volume scheme based on the PROMETHEUS code by
\citet{fryxell1989a} which is an implementation of the ``piecewise
parabolic method'' (PPM) of \citet{colella1984a}.  In order to track
the expanding WD during explosion, a co-expanding uniform grid as in
\citet{roepke2005b}, and \citet{roepke2005c} is used.  Nuclear burning
is included applying a simplified scheme \citep{reinecke2002b,
  fink2010a}: level sets are used to propagate the nuclear burning
flames at the correct speed.  In this thin flame approximation, an
immediate energy release is performed behind the level set
representing the flame surface using the new tables from
\citet{fink2010a} for detonations.  A table for deflagrations was
calculated similar to that work.

All simulations presented here were performed assuming 2D rotational
symmetry on a grid with cylindrical coordinates.  The initial stellar
model was a cold, isothermal ($T=5\times 10^5\,\mathrm{K})$ white
dwarf in hydrostatic equilibrium of mass $M_{WD}=1.401 \, \msun$ and
electron fraction $Y_e = 0.49886$, which corresponds to a central
density of $\rho_c = 2.9 \times 10^9 \gcc$.  For nuclear energy
generation purposes, the composition was assumed to be 50\%
\nuc{16}{O} and 50\% \nuc{12}{C} by mass homogeneously throughout the
star.  The grid resolution was $512 \times 512$ cells for simulations
that were restricted to one hemisphere (assuming mirror symmetry
across the equator), and $512 \times 1024$ cells for simulations that
included both hemispheres respectively.  Three different sets of
explosion models were investigated. They are discussed briefly in turn
below.

\subsection{Centrally ignited pure deflagration}
\label{sec:DEF}
In the pure turbulent deflagration model, the burning was ignited
centrally in a simple spherical shape of radius $150 \, \mathrm{km}$
with a superposed two-period cosine-wave perturbation of amplitude $30
\, \mathrm{km}$ \citep[the C3 ignition configuration as described
by][]{reinecke1999b}.  This setup assumed mirror symmetry across the
equator.  This explosion produced $\sim 0.45 \, \msun$ of iron group
elements (IGE).

\subsection{Centrally ignited delayed detonation}
\label{sec:DDT}
In the second model, an initial deflagration was ignited centrally in
exactly the same C3 configuration as in the pure deflagration runs
(see above).  This time, the deflagration transitioned to a detonation
after the flame had entered the distributed burning regime and the
Karlovitz number\footnote{The use of the Karlovitz number is not
  rigorous here as it depends on the concept of a laminar flame speed
  which does not exist in the distributed burning regime. Nonetheless,
  it is formally used in order to characterize the strength of
  turbulence.} exceeded 250 \citep[see][]{kasen2009a}.  This setup
also assumed mirror symmetry across the equator.  This explosion
produced $\sim 0.57 \, \msun$ of IGE.

\subsection{Multi point ignition delayed detonation}
\label{sec:deldet}
For the third set of explosion simulations, the deflagration was
ignited in 100 ignition kernels of radius $6 \, \mathrm{km}$ randomly
drawn from a Gaussian radial distribution with a standard deviation of
$150 \, \mathrm{km}$ and from a uniform distribution in angle,
resulting in ignition kernels distributed within the inner $306 \,
\mathrm{km}$ of the WD core \citep[corresponding to the ignition model
DD2D\_iso\_06 of][]{kasen2009a}. The detonation was triggered in the
same way as the centrally ignited delayed detonation described
above. This setup simulated both hemispheres independently and
represented a full star in axisymmetry.  This explosion produced $\sim
1.03 \, \msun$ of IGE.

\subsection{Post-processing}
\label{sec:pp}
In the post-processing step, isotopic nucleosynthetic yields for 384
nuclides are calculated by integration of a nuclear reaction network
over the recorded temperature and density time histories along the
paths the tracer particles took in the hydrodynamic simulation.  Apart
from the added feature to allow for variable tracer particle masses
(see section~\ref{sec:placement}), the post-processing scheme is
identical to the one used in \citet{fink2010a}, which in turn is based
on \citet{travaglio2004a}.  Further detailed description of how the
code solves the nuclear network can be found in
\citet{thielemann1990a}, \citet{thielemann1996a}, and
\citet{iwamoto1999a}.  For the reaction rate libraries we use the 2009
release of REACLIB \citep{rauscher2000a}, and for the weak reaction
rates we use \citep{langanke2000a}.  For post-processing purposes, it
was assumed that the initial composition of the unburned material was
a homogeneous mix consisting of 50\% \nuc{16}{O}, 47.5\% \nuc{12}{C},
and 2.5\% \nuc{22}{Ne} by mass (to account for solar metalicity of the
ZAMS progenitor).

\section{Convergence study}
\label{sec:conv}

\begin{figure*}
  \includegraphics[height=8.5in,clip] {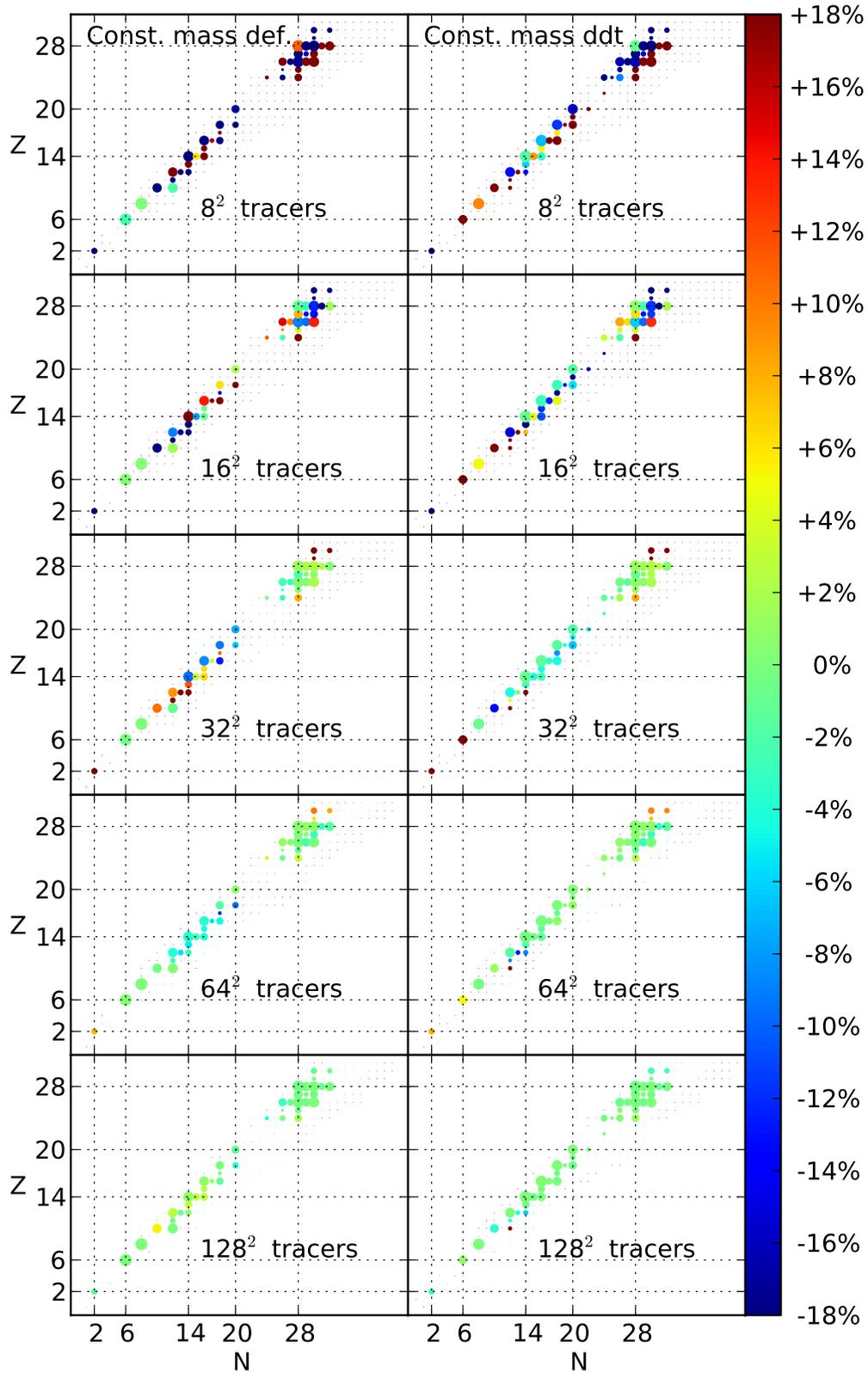}
  \caption{Final ($t=10~\mathrm{s}$) nuclide mass fraction differences
    $\Big[\frac{X_i(N^2)-X_i(256^2)}{X_i(256^2)}\Big]$ in percent
    for a sequence of increasing total tracer particle number compared
    with the highest resolved case containing $256^2$ tracer
    particles.  The underlying hydrodynamical simulation is the same
    central ignition pure deflagration for the left column (see
    Section~\ref{sec:DEF}) and the same centrally ignited delayed
    detonation model for the right column (see
    Section~\ref{sec:DDT}). In all cases the traditional constant
    tracer particle mass approach was used.  The radius, $s_i$, of the
    markers increases with mass fraction $X_i$ according to
    $s_i=\max\{0.1,\,29.9[\log_{10}(X_i)+5]/5+0.1\}$ in arbitrary
    units.}
  \label{fig:mfdiff}
\end{figure*}

\begin{figure*}
  \includegraphics[height=8.5in,clip]{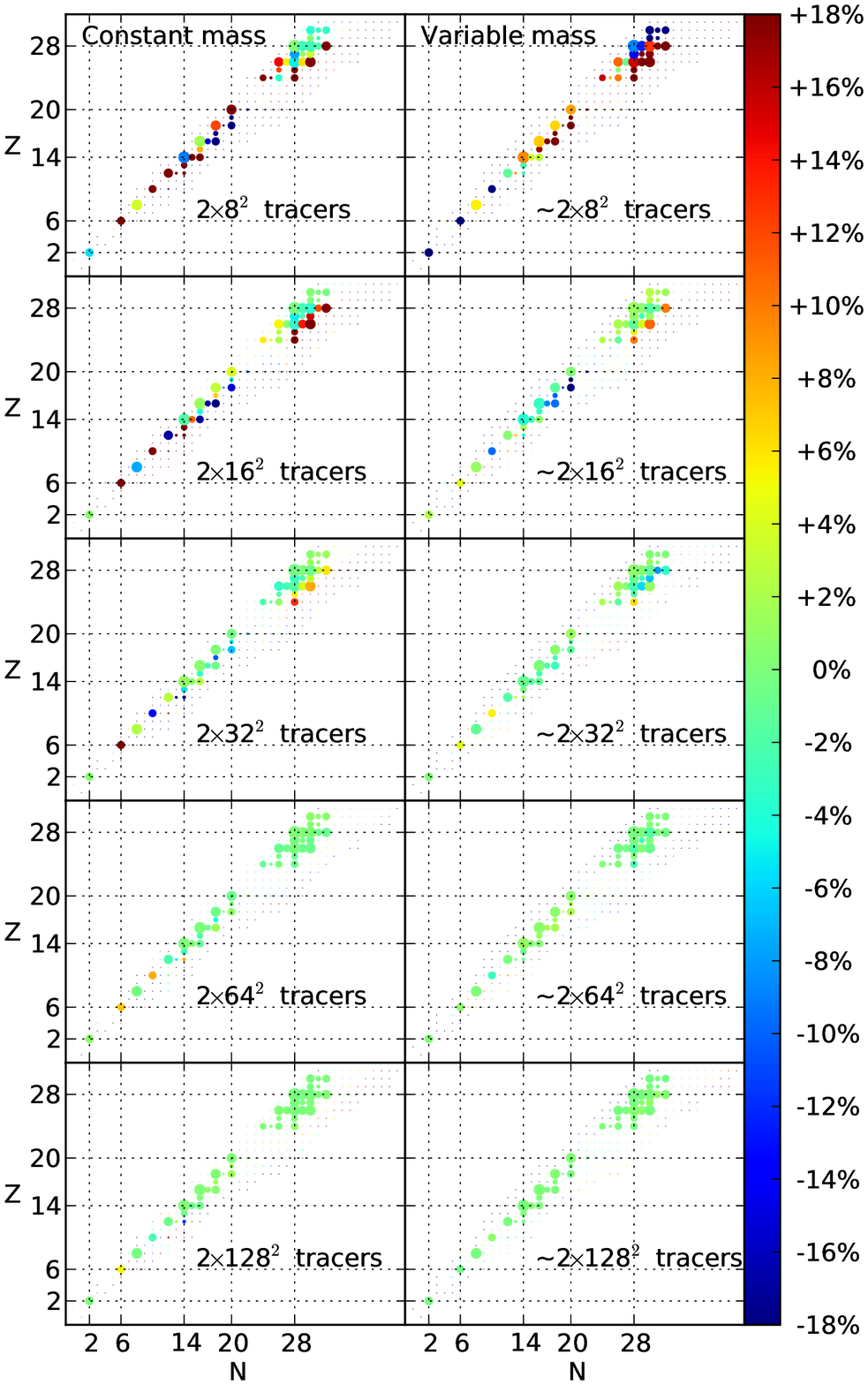}
  \caption{Final ($t=10~\mathrm{s}$) nuclide mass fraction differences
    $\Big[\frac{X_i(2 \times N^2)-X_i(2 \times 256^2)}{X_i(2 \times
      256^2)}\Big]$ in percent for a sequence of increasing total
    tracer particle number compared with the highest resolved case
    containing $2 \times 256^2$ tracer particles.  The underlying
    hydrodynamical simulation is the same multi-spot ignition delayed
    detonation model (see Section~\ref{sec:deldet}) for all cases.
    The left column is for tracer particles of constant mass, whereas
    for the right column the variable tracer mass approach was used
    (see Section~\ref{sec:placement} and Fig.~\ref{fig:vol_mass}).
    The radius, $s_i$, of the markers increases with mass fraction
    $X_i$ according to
    $s_i=\max\{0.1,\,29.9[\log_{10}(X_i)+5]/5+0.1\}$ in arbitrary
    units.}
  \label{fig:deldetdiff}
\end{figure*}

An important question to ask is how many tracer particles are needed
to converge on the final integrated mass fractions after
freeze-out. Only a limited amount of work has been done so far in this
direction.  \citet{travaglio2004b} performed in the context of core
collapse supernova models a 1D resolution study on the number of
tracer particles and conclude that for 2000 zones convergence is
reached at 1000 particles. They stated that the results may not be
applicable in the multi-dimensional case and that a resolution study
for 2D is in preparation, which, however, apparently was not
published.  \citet{brown2005a} post-processed a 2D pure deflagration
explosion with 10000 (constant mass) tracer particles. They found,
that when choosing a random subset of 5000 tracer particles, the
\nuc{56}{Ni} mass changed by 8\% (they didn't state whether it
increased or decreased).  Most recently \citet{roepke2006b},
post-processed 3D pure deflagration models with $27^3$ tracer
particles in the productions runs. They looked into variations of the
\nuc{56}{Ni} mass by increasing the number of tracers to $35^3$ and
concluded that differences were on the percent level.  To date, a more
detailed study spanning a larger range of tracer particle number and
including isotopes different from \nuc{56}{Ni} has not been done.

Here we present a more thorough resolution study in the number of
tracer particles for the three different explosion models outlined in
Section~\ref{sec:sim}.  For the pure deflagration (see
Section~\ref{sec:DEF}) and the centrally ignited delayed detonation
simulations (see Section~\ref{sec:DDT}) we have calculated
nucleosynthetic yields based on 64, 256, 1024, 4096, 16384 and 65536
constant mass tracer particles. For the multi-point ignition delayed
detonation (see Section~\ref{sec:deldet}), which did not assume mirror
symmetry across the equator, we have calculated nucleosynthetic yields
based on 128, 512, 2048, 8192, 32768 and 131072 constant mass tracer
particles.  This corresponds in all cases to 8, 16, 32, 64, 128, and
256 tracer particles per axis and direction. In addition, we have
calculated yields for the same multi-point ignition delayed detonation
model with variable tracer masses using $M_1=1.05 \, \msun$ and
$M_2=1.355 \, \msun$ as introduced in Section~\ref{sec:placement}.
For the variable tracer mass case the total numbers of tracers were
slightly different\footnote{Small departures from exact powers of 2
  were necessary to assure mass conservation and the constraint that
  tracer particles in the intermediate region all represent the same
  volume.}, i.e. 122, 499, 2024, 8140, 32664, and 130865.  For
practical purposes these numbers can be considered to be the same
resolution.

In this paper we are less concerned with the exact magnitude of the
mass fraction of a given nuclide, but rather with how well the yields
of the run that included the most tracer particles are reproduced by a
run with fewer tracer particles and how quickly the various mass
fractions converge with total tracer particle number.

In Figs.~\ref{fig:mfdiff} and \ref{fig:deldetdiff} we show, for
sequences of runs with increasing tracer particle numbers, percent
differences of nuclide mass fractions compared to the mass fractions
obtained in the run that included the highest number of tracer
particles, which serve as ``reference values''. To visually emphasize
nuclides with larger mass fractions, the symbol radius, $s_i$,
increases logarithmically from $s_i = 0.1$ for nuclides with mass
fraction $X_i \leq 10^{-5}$ to a possible maximum of $s_i = 30$ in
arbitrary units.

The left column of Fig.~\ref{fig:mfdiff} shows results for the pure
deflagration setup, which used tracer particles of constant mass (see
Section~\ref{sec:DEF}).  The two cases containing the least number of
tracer particles, $8^2$ and $16^2$, show globally poor agreement of
the nuclide mass fractions with the reference values. The case
containing $32^2$ particles already shows good agreement of most
Fe-peak nuclei, but many of the most abundant intermediate mass nuclei
are under-produced on the $5\%$-level. Agreement with the reference
values is only slightly improved by going to $64^2$ tracer particles.
Using $128^2$ tracer particles finally reproduces the reference values
of the most abundant nuclei globally on the $2\%$-level, with a few
exceptions, most notably isotopes of Ne, Mg, and Al.

The right column of Fig.~\ref{fig:mfdiff} shows results for the
centrally ignited delayed detonation setup, which also used tracer
particles of constant mass (see Section~\ref{sec:DDT}).  Similar to
the pure deflagration case, the two low resolution cases show globally
poor agreement of the nuclide mass fractions with the reference
values. Already the $32^2$ tracer particle case shows rather good
agreement with the high resolution reference run for the majority of
the more abundant Fe-peak and intermediate mass nuclides.  Compared to
the pure deflagration, this delayed detonation agrees better with its
references values for intermediate tracer particle numbers
(i.e. $32^2$ and $64^2$). This can be traced back to the fact that the
delayed detonation model synthesizes nuclides such as \nuc{28}{Si},
\nuc{32}{S}, \nuc{36}{Ar} or \nuc{40}{Ca} copiously during the
detonation phase in a large volume where even the constant tracer
particle mass implementation results in a tracer particle number
density high enough to adequately sample the morphology.  Notable
exceptions to the better agreement are nuclides that are predominantly
synthesized in the outer layers of the star during the detonation
phase, such as e.g. \nuc{20}{Ne}, \nuc{26}{Mg} or \nuc{27}{Al} (see
the discussion of the multi-spot ignition delayed detonation model
further below).

\begin{figure}
  \includegraphics[width=\columnwidth]{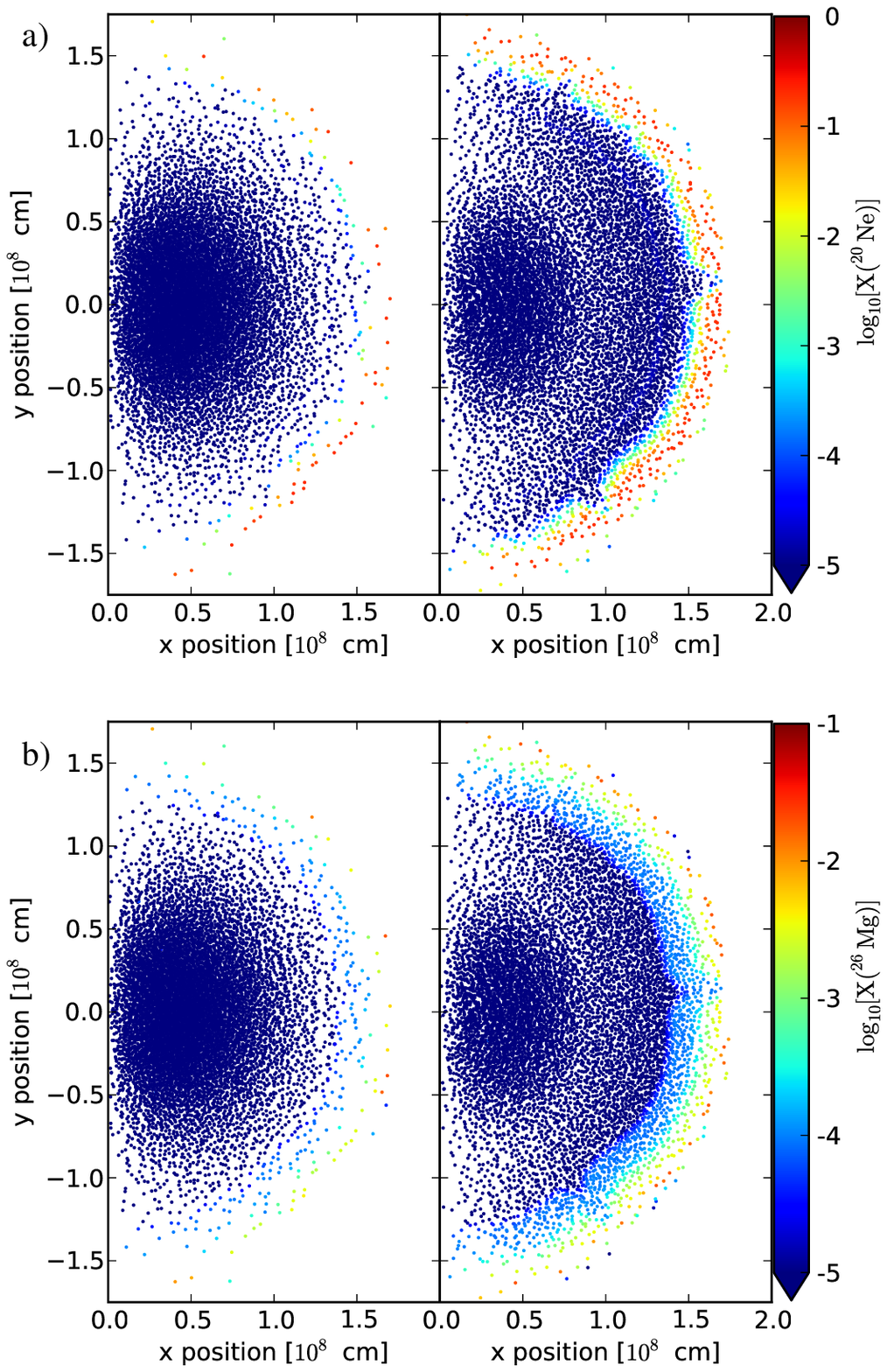}
  \caption{Initial spatial distribution of constant (left, $N=8192$)
    and variable (right, $N=8140$) mass tracer particles colored by
    the final mass fraction of \nuc{20}{Ne} (top panel) and
    \nuc{26}{Mg} (bottom panel) after freeze-out ($t=10~\mathrm{s}$).}
   \label{fig:ne20mg26}
\end{figure}

\begin{figure}
  \includegraphics[width=\columnwidth]{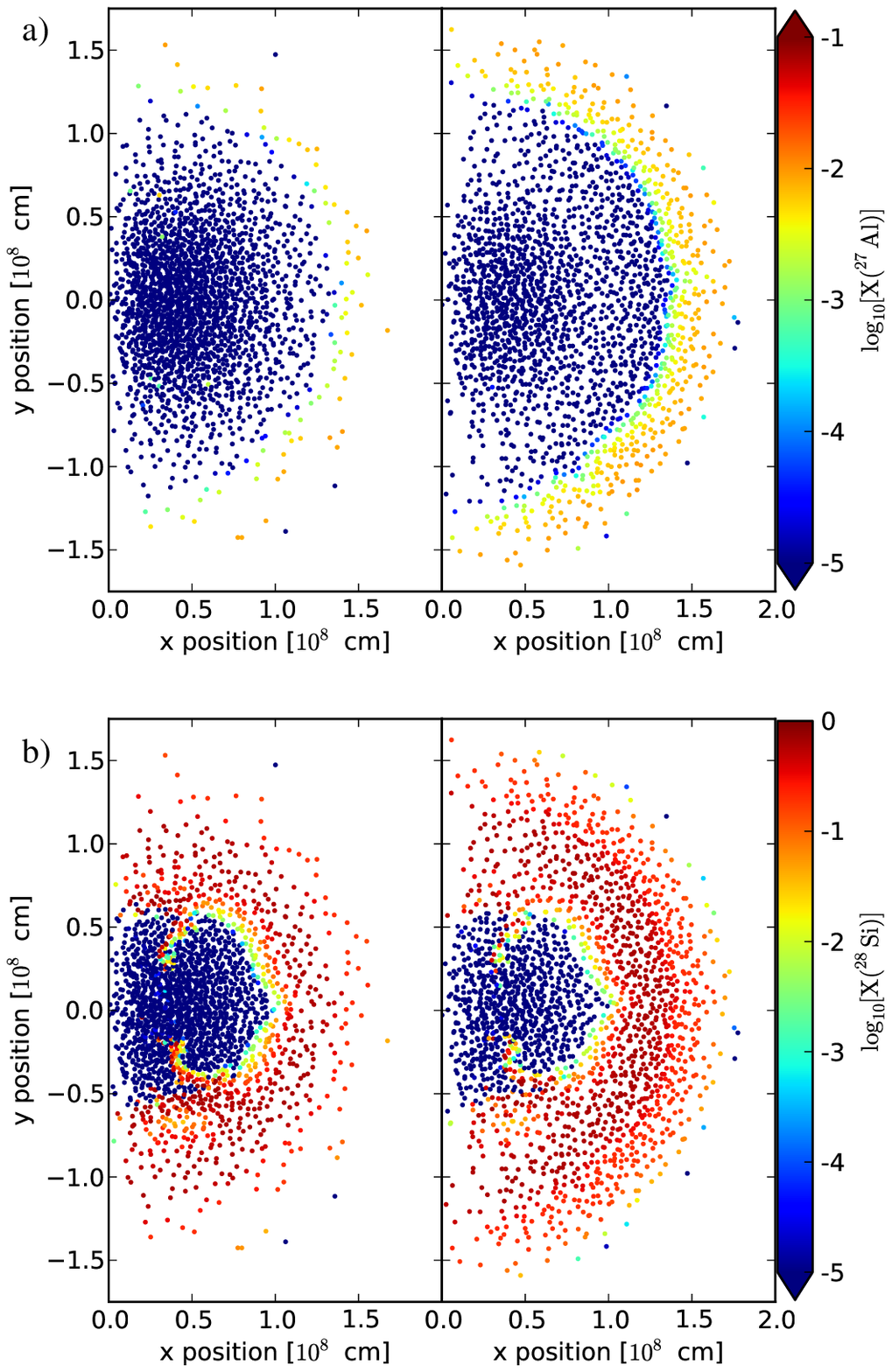}
  \caption{Initial spatial distribution of constant (left, $N=2048$)
    and variable (right, $N=2024$) mass tracer particles colored by
    the final mass fraction of \nuc{27}{Al} (top panel) and
    \nuc{28}{Si} (bottom panel) after freeze-out ($t=10~\mathrm{s}$).}
   \label{fig:al27si28}
\end{figure}

Fig.~\ref{fig:deldetdiff} shows results for the multi-spot ignition
delayed detonation setup (see Section~\ref{sec:DEF}), with constant
tracer particle mass runs in the left column, and variable tracer
particle mass runs (see Section~\ref{sec:placement}) in the right
column. In spite of the differences in explosion model and
nucleosynthetic yields, the sequence of the constant tracer mass case
is qualitatively very similar to the centrally ignited delayed
detonation (compare right column of Fig.~\ref{fig:mfdiff} to the left
column of Fig.~\ref{fig:deldetdiff}).  In all but the lowest
resolution case, using a variable tracer mass approach results in a
global improvement of the agreement with the reference values and
improved convergence characteristics (compare right and left columns
of Fig.~\ref{fig:deldetdiff}).  This improvement is due to the fact
that the variable tracer mass approach alleviates the problem of low
tracer particle density in the outer layers of the star inherent to
the constant mass approach. Abundance gradients of problematic
nuclides such as e.g. \nuc{20}{Ne} (see Fig.~\ref{fig:ne20mg26}a),
\nuc{26}{Mg} (see Fig.~\ref{fig:ne20mg26}b) or \nuc{27}{Al} (see
Fig.~\ref{fig:al27si28}a), which are produced in an incomplete burn
during the detonation phase at low density in the outer layers of the
star, are much better resolved with the variable tracer mass
implementation (compare right and left panels of
Figs.~\ref{fig:ne20mg26}a, \ref{fig:ne20mg26}b) and
\ref{fig:al27si28}a.  For nuclides which qualitatively agree equally
well with their reference values for both constant and variable tracer
mass cases, such as e.g. \nuc{28}{Si}, the variable tracer mass
implementation results in a much improved spatial resolution of the
abundance distribution morphology in the outer regions at the cost of
a slightly worse resolution in the inner regions (see
Fig.~\ref{fig:al27si28}b).

\section{Summary and conclusions}
\label{sec:conc}
We have performed a resolution study in the number of tracer particles
required to converge on the integrated nucleosynthetic yield
calculations for three different SN Ia explosion models.  We find that
for all explosion models, total nucleosynthetic yields appear to have
converged (with few exceptions, see Section~\ref{sec:conv}) for
abundant nuclides with mass fractions $> 10^{-5}$ to better than $2\%$
at 128 tracer particles per axis and direction.  Agreement at better
than the $5\%$ level is already achieved for many of the most abundant
nuclides at particle numbers as low as 32 per axis and direction.

Based on these results, we extrapolate that published yields from the
literature
\citep[][]{travaglio2004a,travaglio2005a,brown2005a,roepke2006b,maeda2010a}
are based on sufficiently high numbers of tracer particles per axis
that an adequate to good prediction of the most abundant nuclei is
given.  Furthermore, we have shown that for isotopes whose origin is
near the surface of the WD, such as e.g. \nuc{20}{Ne}, \nuc{26}{Mg} or
\nuc{27}{Al}, better convergence can be achieved if the constraint of
constant tracer mass is relaxed and a choice for the tracer particle
masses is made that better spatially resolves the outer layers.

The question of how many particles are required to get converged
spectra and light curves from subsequent radiative transfer
calculations is related.  It seems plausible that a convergence of the
total integrated yields is a necessary but not sufficient constraint.
For the radiative transfer not only the total mass of a given isotope
is important but also its location in velocity space.  It seems that
the method of variable mass tracer particles is also promising in this
context.  Especially simulations where the nucleosynthesis is
dominated by a detonation should greatly benefit from a variable
tracer mass distribution, since the mass fractions in the inner
regions vary smoothly radially and therefore a de-refinement in the
inner part is not a high price to pay for increased resolution in the
outer part.  However, the reconstruction of an isotopic abundance for
every computational cell of the radiative transfer calculation from
the tracer particle distribution is not unique.  The resultant spectra
and light curves depend somewhat on particular choice of
reconstruction algorithm used. These questions will likely be
addressed in a future study.

\section*{Acknowledgements}
We want to thank Ashley Ruiter, Keiichi Maeda, Stuart Sim, Markus
Kromer and Wolfgang Hillebrandt for helpful discussions and comments.
The simulations presented here were carried out at the Computer Center
of the Max Planck Society, Garching, Germany.  This work was supported
by the Deutsche Forschungsgemeinschaft via Emmy Noether Program (RO
3676/1-1) and the Excellence Cluster EXC~153.

\bibliographystyle{../mn2e} 
\bibliography{../astrofritz.bib}

\end{document}